# Automated metrics calculation in a dynamic heterogeneous environment

Craig Boucher, Ulf Knoblich, Daniel Miller, Sasha Patotski, Amin Saied, Venky Venkateshaiah (Microsoft Corp.)

## 1   Introduction

A consistent theme in software experimentation at Microsoft has been solving problems of experimentation at scale for a diverse set of products. Running experiments at scale (i.e., many experiments on many users) has become state of the art across the industry [1]. However, providing a single platform that allows software experimentation in a highly heterogenous and constantly evolving ecosystem remains a challenge. In our case, heterogeneity spans multiple dimensions. First, we need to support experimentation for many types of products: websites, search engines, mobile apps, operating systems, cloud services and others. Second, due to the diversity of the products and teams using our platform, it needs to be flexible enough to analyze data in multiple compute fabrics (e.g. Spark, Azure Data Explorer), with a way to easily add support for new fabrics if needed. Third, one of the main factors in facilitating growth of experimentation culture in an organization is to democratize metric definition and analysis processes. To achieve that, our system needs to be simple enough to be used not only by data scientists, but also engineers, product managers and sales teams. Finally, different personas might need to use the platform for different types of analyses, e.g. dashboards or experiment analysis, and the platform should be flexible enough to accommodate that. This paper presents our solution to the problems of heterogeneity listed above.

## 2   Common challenges in experimentation at scale

Being able to run many A/B experiments concurrently with many users has been a well-understood problem, and many robust solutions already exist [1,2]. However, analyzing the results of the experiments and deriving value for the product is still difficult to do at scale. Even within a single product there can be a lot of diversity in terms of the moving components that the owners might want to improve via A/B testing: there could be desktop, web and mobile versions, each of them with different UIs, connecting to various backend services, using various ML models. Naturally, there will be multiple teams working on these different aspects of the product. Such diversity exacerbates many of the common challenges associated with experimentation in modern product development.

In a cross-team effort spanning multiple years, our team built a cross-platform code generation framework for trustworthy metric computation. In the past, experiment analysis was tightly coupled to a *single data source*. Data scientists would *manually* write queries in a *fabric-specific language* to compute the metrics. This made it difficult to scale the number of metrics computed and made migrating analyses to a different compute fabric an extremely laborious process. It was hard to ensure the *trustworthiness of the statistical analysis* for all these metrics. It was cumbersome to create *focused analyses* that would only include metrics for a specific part of the product, and it was difficult to *change the collection of segments* for a given query. It required significant engineering resources to add *advanced new statistical features* like variance reduction [4], and there was no reliable way to *share metric definitions* between queries, let alone re-use a metric for both experimentation and *in a business dashboard*. Our solution addresses these pain points in several ways:

1) To have a comprehensive understanding of users' experience with the product, many of our partner teams have hundreds (in some cases even thousands) of metrics monitoring different aspects of the product/business. Our team has developed a **domain-specific language** internally called MDL (Metrics Definition Language) for generic metric definitions, which can reduce dozens of lines of SQL code to a single line metric definition (see examples below). Simultaneously, this makes metrics easier to read and write, and also reduces the surface for errors.
2) In our system, the definitions of all the metrics are stored in a single repository. This allows multiple people to coordinate their contributions to the metric set and allows **queries to be shared** between analyses and experiments. Because metrics to compute in each analysis are automatically selected, having 10k metrics in a single place does not affect the performance of the analysis computations.
3) Particularly once there is a large set of metrics, it is often desirable to conduct focused analyses that only include a **subset of all metrics** concentrating on a particular aspect of the product or feature. This simplifies users' understanding of the analyses results, as well as improves the compute efficiency. Users can create metric groups and assemble scorecard layouts based on these groups, allowing easily configurable analyses that are tailored to the needs and goals of each team.
4) When analyzing the results of a controlled experiment, it is not enough to compute metric values in treatment and control populations. You also need **statistical tests** to understand the treatment effect. For a single metric, the process of choosing the correct variance estimate based on the characteristics of the metric is not trivial and error-prone; having to do this manually for each metric is cumbersome and very likely to result in incorrect results. The correct statistical test for each metric is automatically selected and the necessary computations for calculating those tests added. This means that no deep statistical knowledge is required for defining metrics and adding them to a scorecard, making it possible for these tasks to be owned and completed by the feature teams, improving agility and adoption.
5) In addition to statistical analysis, it is crucial to be able to investigate different **segments of the experiment population**. For example, it often happens that users in different countries react differently to the same UI change. A website change can work (or not work) differently in different browsers. The product owner should be able to quickly configure which segments (or slices) of the data to look into, and should be able to easily combine several of them, e.g. look at the data from users in a specific country *and* using a specific browser. All the logic necessary to analyze these segments is automatically added, the user only needs to supply the list of segments they want to compute. These segments are defined once for a given product and are reused between experiments and analyses. Moreover, the user can define stable groups of segments to reuse between experiments. For example, one can have "client experiment segments", "UI change segments" and so on.
6) Data can live in different frameworks (e.g., Hadoop, Spark, Cosmos DB) each supporting their own syntax/language for defining metrics. It is commonplace for metric definitions to be tightly coupled to this framework (e.g. if metric definitions are hard-coded in SQL), meaning that a change in framework involves costly and error-prone re-authoring of metrics. We have decoupled metric definition and metric computation so that one metric can be computed in **many frameworks** without being re-written. This also allows for scenarios such as having data in a transient near-real-time system that can be used to generate fast scorecards after hours or even minutes, while using a slower, more persistent system for scorecards spanning several weeks, all using the same metric definitions. Our architecture is extensible in that support for new frameworks can easily be incorporated.

7) Decoupling metric definitions from the analysis code means that bug fixes or implementation of new advanced features like variance reduction [4], finding the "best" randomization seed [5, 6] or running a "retrospective" A/A test [6] are comparatively easy. The code only has to be written once, can be validated on the existing use cases, and is immediately available for all metric sets and scorecards.

8) Traditionally there is a distinction between experimentation metrics and business metrics (e.g., used for **reporting**). In our system, metrics are context agnostic, meaning that the same metric can be used in a business or experimentation context. In the experimentation context, control and treatment metrics as well as the necessary statistics required for analysis are generated. For business reporting, dashboards, etc., these can be omitted, making the resulting computations faster and cheaper.

## 3 System Overview

Having highlighted some of the high-level features, let us briefly describe a typical workflow from a user's perspective to motivate our system architecture. First, a user defines a collection of metrics called a metric set. These metrics are designed for a specific product and its underlying telemetry data but can be reused across experiments and analyses. This usually first happens before an experiment is run (it is considered good practice to think about the metrics you want to measure before the experiment), but metric definitions can be added or edited at any time and retro-actively applied to previous experiments, as well. Once an experiment is running and data is available, a configuration object is used to instantiate a specific analysis task. This configuration is often sent by the experiment management system automatically scheduling pre-defined analyses but can also be a one-off request from a user. The provided information is used to create an abstract logical model of the computation called **metrics plan** which is then altered through a series of transformations, and converted into code for the requested compute fabric. This code is then submitted for computation and the results are displayed to the user. To learn more about the overall structure of our experimentation platform see [2].

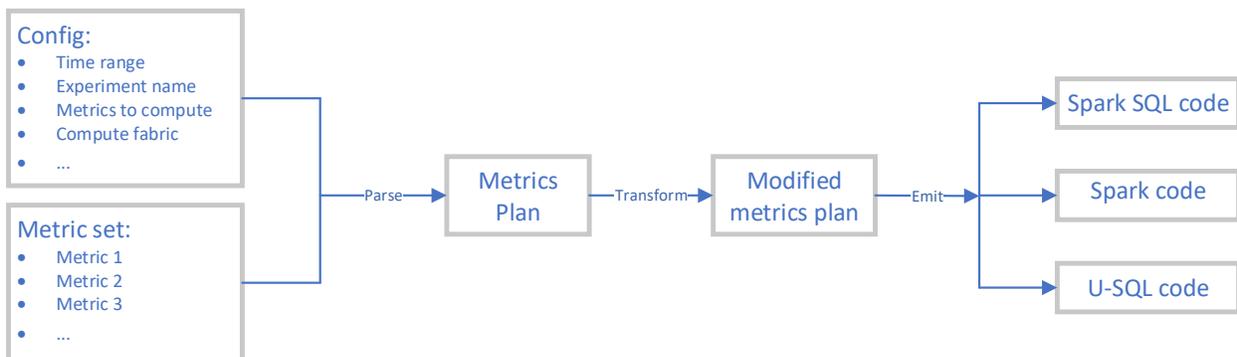

*Figure 1 -- High level system diagram*

## 3.1 Metric sets and Metrics Definition Language

A metric set is a collection of all possible metrics defined for a given product. The metrics are defined in MDL, which is a metric definition language built internally by Microsoft's AI Platform team to enable authoring experimentation metrics efficiently and accurately by a wide range of users. MDL is fabric agnostic, with logical definitions that allow customers to define their metrics once and reuse them across multiple experiments, analyses and data fabrics.

As an example, suppose we are defining metrics for experimentation on a website. When a user browses the website, each page they visit gets logged in a single row of data, together with the actions the user performed on that page and some information about the page, such as page load time. If any of those actions was a purchase, we log the corresponding revenue in the Revenue column. Using MDL, we can very succinctly define metrics, for example:

- "average revenue per user" could be written as: `Avg(Sum<User>(Revenue))`. Here we first sum all the values of the Revenue column for each user to get the total user's revenue across all of that user's pages, and then compute the average of these values across all users.
- "95th percentile of page load time" could be defined as: `Percentile(PageLoadTime, 95)`.

## 3.2 Metrics plan

The metrics plan is the abstraction layer separating the core logic of metrics computation from specific compute fabrics. After consuming the metric definitions authored using MDL and a config file, the first step is to parse them into a metrics plan, a directed acyclic graph containing syntax trees of expressions. Metrics plans are fabric agnostic and completely describe the requested computation. Moreover, they can be modified to add expressions (e.g. for variance calculation, see below), remove or replace expressions, etc. In this way we decouple the *description* of the computation from the actual *implementation* which depends on the compute fabric. Such decoupling makes is very easy to add support for new compute fabrics, since most of the logic is not fabric specific. This makes the system more agile and future-proof.

### 3.2.1 Transformations

After parsing the metric set and configuration into a metrics plan, a series of transformations is performed. First, the metrics plan is normalized into a canonical form. Second, we perform enrichments such as adding expressions for variance computation and segmentation. Lastly, before emitting the code, we perform fabric-agnostic optimizations (e.g. deduplicating expressions) as well fabric-specific ones.

#### 3.2.1.1 Variance

A fundamental step when analyzing an experiment is to compute the variance of the given metrics. By analyzing the graph structure of the metrics plan, the appropriate variance formula for a given metric can be determined. The standard variance formula is used when the metric is based on an independent sample, e.g. a per-user metric in an experiment where randomization was user-based. The "average revenue per user" metric defined above is an example of such a metric. The delta method [3] is used when the metric is based on a potentially dependent sample, e.g. a per-page metric in an experiment where the randomization is done per user. The "page load time" metric above is an example of such a metric.

There are several advantages to this approach. First, this frees the user from having to define the variance for each metric explicitly, making it more user-friendly and less error prone. By automatically selecting the appropriate variance to compute for each metric the analysis is also more trustworthy. Finally, the logic that powers the variance computation has been abstracted away from the compute-fabric making it reusable and providing a more consistent analysis.

### 3.2.1.2   Optimization

Before emitting the code, a series of optimization transformations is applied to the metrics plan to improve the final script's performance. For example, we can remove any expressions that are not needed for the metrics to be computed, we can alias frequently used expressions to deduplicate the computations, we can remove unnecessary null checks and comparisons, and others.

### 3.2.2   Code Emission

Since all the complicated logic has been applied to the fabric-agnostic metrics plan at the previous step, the code generation part is relatively straightforward. It is defined by a mapping from nodes in the syntax tree to the corresponding representation in the chosen compute fabric's language. Adding support for a new compute fabric only requires defining how to implement generic computation constructs (e.g., join, sum) in the syntax for that fabric.

# 4   References

[1] R. Kohavi, A. Deng, B. Frasca, T. Walker, Y. Xu, N. Pohlman. 2013. *Online Controlled Experiments at Large Scale*. Proceedings of the 19th ACM SIGKDD International Conference on Knowledge Discovery and Data Mining. KDD 2013, pp. 1168 – 1176, 2013, Chicago, Illinois, USA.

[2] P. Dmitriev, L. Ulanova, S. Gupta, S. Bhardwaj, A. Fabijan, P. Raff. 2018. *The Anatomy of a Large-Scale Online Experimentation Platform*. 10.1109/ICSA.2018.00009.

[3] A. Deng, U. Knoblich, J. Lu.  2018. *Applying the Delta Method in Metric Analytics: A Practical Guide with Novel Ideas*. Applied Data Science Track Paper. KDD 2018, pp. 233 – 242, August 19-23, 2018, London, United Kingdom.

[4] A. Deng, Ya Xu, R. Kohavi, T. Walker. 2013. *Improving the Sensitivity of Online Controlled Experiments by Utilizing Pre-Experiment Data*. WSDM 2013, pp. 123 – 132, February 4-8, 2013, Rome, Italy.

[5] X. Li, P. Ding. 2019. *Rerandomization and Regression Adjustment*. Preprint, https://arxiv.org/abs/1906.11291

[6] R. Kohavi, A. Deng, B. Frasca, R. Longbotham, T. Walker, Ya Xu 2012. *Trustworthy online controlled experiments: five puzzling outcomes explained*. Applied Data Science Track Paper. KDD 2012, pp. 786 – 794, August 12-26, 2012, Beijing, China.